\documentclass{elsart}
\usepackage{epsfig}
\begin{document}        
\begin{frontmatter}    
\title{\bf Spherical Neutral Detector for VEPP-2M collider}   
\author{ M.N.Achasov\thanksref{adrr},}    
\author{V.M.Aulchenko,} 
\author{S.E.Baru,} 
\author{K.I.Beloborodov,}
\author{A.V.Berdyugin,} 
\author{A.G.Bogdanchikov,} 
\author{A.V.Bozhenok,} 
\author{A.D.Bukin,} 
\author{D.A.Bukin,}
\author{S.V.Burdin,} 
\author{T.V.Dimova,} 
\author{S.I.Dolinsky,} 
\author{A.A.Drozdetsky,} 
\author{V.P.Druzhinin,}
\author{M.S.Dubrovin,} 
\author{I.A.Gaponenko,} 
\author{V.B.Golubev,}
\author{V.N.Ivanchenko,} 
\author{A.A.Korol,} 
\author{S.V.Koshuba,}
\author{G.A.Kukartsev,}
\author{E.V.Pakhtusova,} 
\author{V.M.Popov,}
\author{A.A.Salnikov,} 
\author{S.I.Serednyakov,}
\author{V.V.Shary,}
\author{V.A.Sidorov,} 
\author{Z.K.Silagadze,}
\author{Yu.V.Usov,} 
\author{A.V.Vasiljev,} 
\author{Yu.S.Velikzhanin}
\author{A.C.Zakharov}
	
\thanks[adrr]{ E-mail: achasov@inp.nsk.su, FAX: +7(383-2)34-21-63}          
\address{ G.I.Budker Institute of Nuclear Physics, 
          Siberian Branch of the Russian Academy of Sciences 
	  and Novosibirsk State University,
          Novosibirsk,
          630090,   
          Russia }        
\date{}    
\begin{abstract}
 The Spherical Neutral Detector (SND) operates at VEPP-2M collider in
 Novosibirsk studying $e^+e^-$ annihilation in the energy range up
 to 1.4 GeV. Detector consists of a fine granulated spherical 
 scintillation calorimeter with 1632
 NaI(Tl) crystals, two cylindrical drift chambers with 10 layers of sense
 wires, and a muon system
 made of streamer tubes and plastic scintillation counters. The detector
 design, performance, data acquisition and processing are described.
\end{abstract}
\end{frontmatter}

\section{Introduction}

 For more than 25 years the VEPP-2M $e^+e^-$ collider has been operating in
 BINP, Novosibirsk, in the center-of-mass energy range $2E_0=0.36 \div 1.4$ GeV
 \cite{vepp2}. Until now its maximum luminosity of
 $L=3 \cdot 10^{30}~\mathrm{cm}^{-2}\mathrm{s}^{-1}$ at $E_0=510$ MeV
 was a record for the machines of this class.
 During this period several
 generations of detectors performed experiments at VEPP-2M. Much of the
 current data on particles properties \cite{pdg} at low energy region
 were obtained in these experiments.
 
 The SND detector is an advanced version of its predecessor
 -- the Neutral Detector
 (ND) \cite{ndnim,ndnm}, which completed its five-year
 experimental program in 1987 \cite{nd}.
 The SND detector \cite{snddaf} operates at VEPP-2M since 1995
 \cite{snd1,snd2,snd3,snd4}. Its main part is a three-layer spherical
 calorimeter consisting of 1632 crystals NaI(Tl). The SND is a general
 purpose detector optimized  for multi-photon final states.
 
\section{General overview}

\begin{figure}
\begin{center}
\epsfig{figure=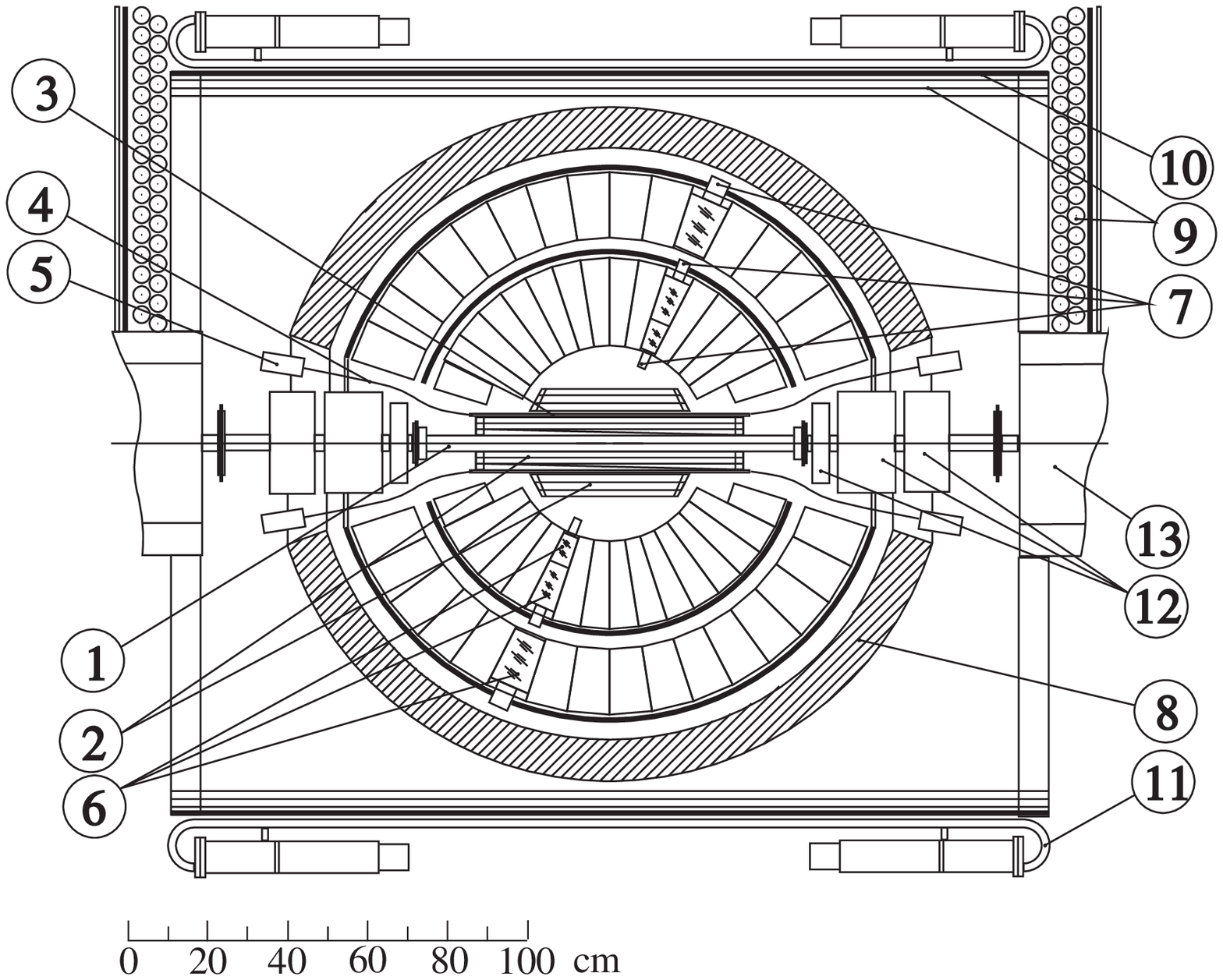,height=7cm}
\caption{SND detector, section along the beams: (1) beam pipe,
         (2) drift chambers, (3) scintillation counter, (4) light guides,
	 (5) PMTs, (6) NaI(Tl) crystals, (7) vacuum phototriodes,
	 (8) iron absorber, (9) streamer tubes, (10) 1 cm iron plates,
         (11) scintillation counters, (12) and (13) elements of collider
         magnetic system.}
\label{sndt}
\epsfig{figure=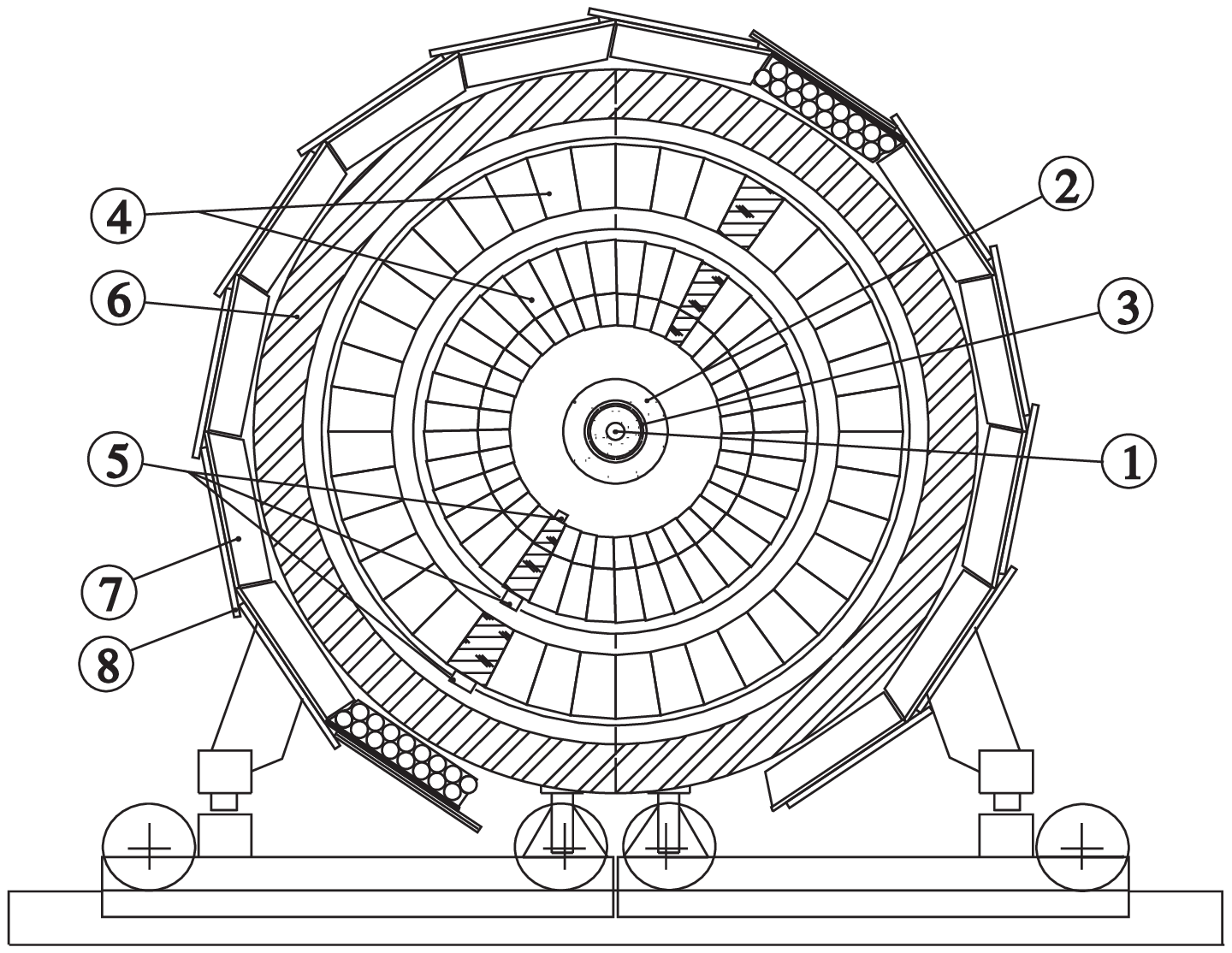,height=7cm}
\caption{SND detector, section across the beams: (1) beam pipe,
         (2) drift chambers, (3) scintillation counter, (4) NaI(Tl)
	 crystals, (5) vacuum phototriodes, (6) iron absorber,
         (7) streamer tubes, (8) scintillation system.}
\label{sndf}
\end{center}
\end{figure}

 The detector layout is shown in Figs.~\ref{sndt},~\ref{sndf}.
 Electron and positron beams collide inside the beryllium beam pipe
 with a diameter of 2 cm and 1 mm wall. The beam pipe is surrounded by
 tracking system consisting of two drift chambers and a cylindrical
 scintillation counter between them. The solid angle coverage of the tracking
 system is about $98\%$ of $4\pi$.

 The three-layer spherical electromagnetic calorimeter based on NaI(Tl)
 crystals surrounds the tracking system. The total calorimeter thickness for
 particles originating from the interaction region is 34.7 cm
 (13.4 $X_0$) of NaI(Tl) and the total solid angle is $90\%$ of $4\pi$.

 Outside the calorimeter a 12 cm thick iron absorber is placed in order to
 attenuate the residuals of electromagnetic showers. It is surrounded by
 segmented muon system which provides  both muon identification and
 cosmic background suppression . Each segment
 consists of two layers of streamer tubes and a plastic scintillation
 counter, separated from the tubes by 1 cm iron plate.
 The iron layer between the tubes and the counter reduces the
 probability of their simultaneous firing by photons produced in
 $e^+e^-$ collisions to less than $1\%$ for 700 MeV photons.

\section{Calorimeter}

\subsection{Calorimeter layout}

\begin{figure}
\begin{center}
\epsfig{figure=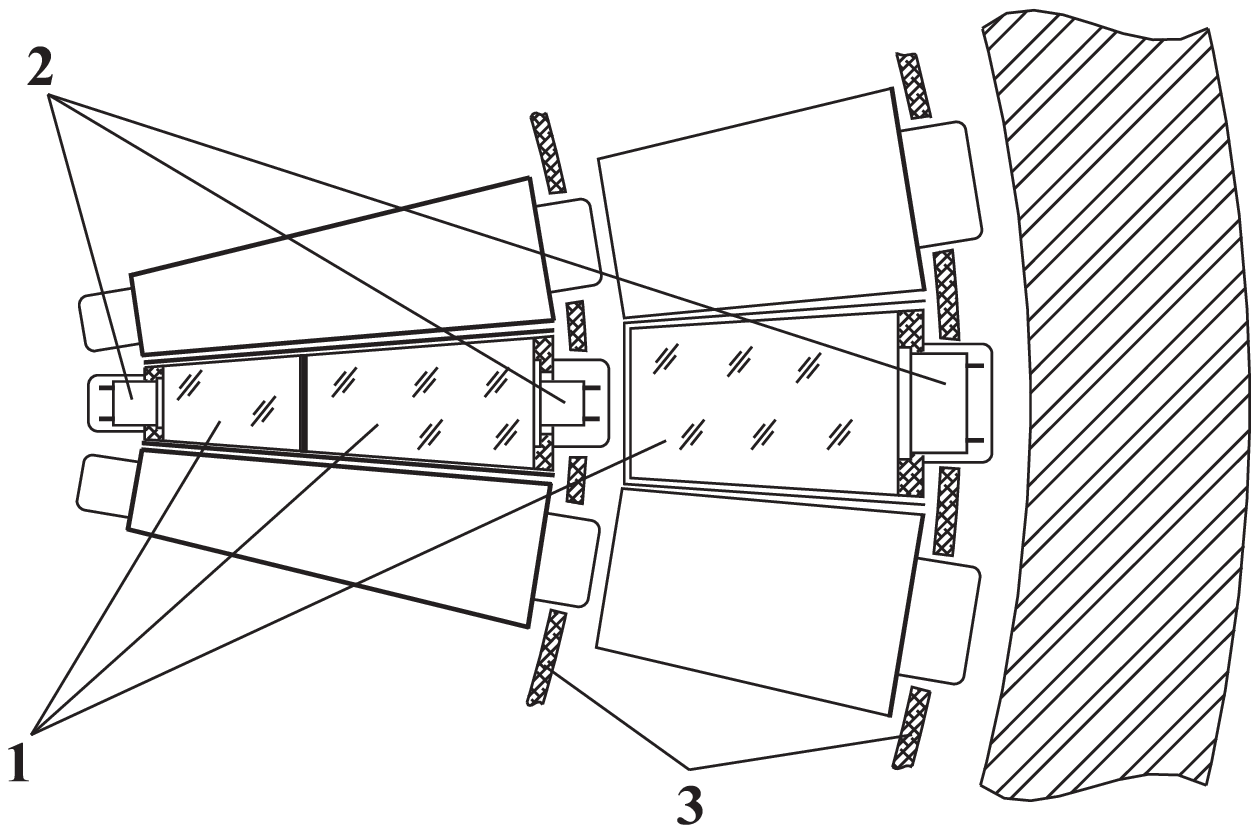,height=7cm}
\caption{NaI(Tl) crystals layout inside the calorimeter: (1) NaI(Tl) crystals,
         (2) vacuum phototriodes, (3) aluminum supporting hemispheres.}
\label{cryst}
\epsfig{figure=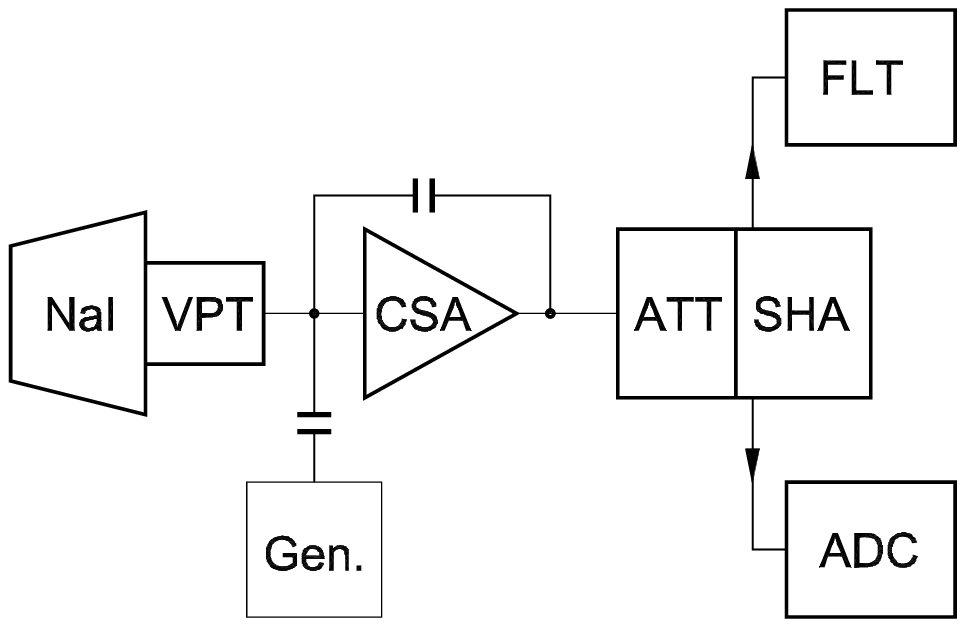,height=7cm,width=20cm}
\caption{Electronics channel of the SND calorimeter: (NaI) NaI(Tl)
         scintillator, (VPT) vacuum phototriode, (CSA) charge-sensitive
	 preamplifier, (ADC) analog to digital converter, (Gen) calibration
         generator, (SHA) shaping amplifier, (ATT) computer-controlled
         attenuator, (FLT) first-level trigger}
\label{chan}
\end{center}
\end{figure}

 Spherical shape of the SND calorimeter provides relative uniformity of
 response over the whole solid angle. Pairs of counters of the two inner
 layers with thickness of 2.9 and 4.8 $X_0$ ($X_0=2.6$~cm) are sealed in thin
 (0.1 mm) aluminum containers, fixed to an aluminum supporting hemisphere
 (Fig.~\ref{cryst}). Behind it, the third layer of NaI(Tl) crystals, 5.7 $X_0$
 thick, is placed. The gap between the adjacent crystals of one layer is
 about 0.5 mm. The total number of counters is 1632, the number of crystals
 per layer varies from 520 to 560. The total mass of NaI(Tl) is 3.5 t. 

 The polar angle coverage of the calorimeter is
 $18^\circ \leq \theta \leq 162^\circ$. The calorimeter is divided
 into two parts:  ``large'' angles $36^\circ \leq \theta \leq 144^\circ$ and
 ``small'' angles --- the rest.
 The angular dimensions of crystals
 are $\Delta\phi = \Delta\theta=9^\circ$ at ``large'' angles and
 $\Delta\phi = 18^\circ, \Delta\theta=9^\circ$ at ``small'' angles. Each layer
 of the calorimeter consists of crystals of eight different shapes.

 The crystal widths approximately match the transverse size
 of an electromagnetic shower in NaI(Tl). Two showers can be distinguished
 if the angle between them is larger than $9^\circ$. If this angle
 exceeds $18^\circ$ the energies of the showers can be measured with
 the same accuracy as for isolated shower. A high granularity of the
 calorimeter is especially useful for the detection of multi-particle events.
 For example, the detection efficiency for 7-photon events of the process
 $\phi \to \eta \gamma$, $\eta \to 3 \pi^0$ is close to $15\%$.

 The light collection efficiency varies from $7\%$ to $15\%$ for crystals of
 different calorimeter layers. The scintillation light signals from the
 crystals are detected by
 vacuum phototriodes \cite{VPT} with a photocathode diameter of 17 mm in the
 first two layers and 42 mm in the third layer. The average photocathode
 quantum efficiency is about $15\%$ and the mean tube gain is
 about 10.
 
 The electronics of the calorimeter (Fig.~\ref{chan}) consists of
\begin{enumerate}
\item
 the charge sensitive preamplifiers (CSA) with a conversion coefficient
 of~0.7~V/pC,
\item
 12-channel shaping amplifiers (SHA) with a remote controlled gain that
 can be set to any value in the range from 0 to a maximum with a resolution
 of 1/255.
\item
 24-channel 12-bit analog to digital converter (ADC) with a maximum
 input signal $U_{\mathrm{max}}=2 \mathrm{V}$,
\end{enumerate}

 Each  calorimeter channel can be tested using
 a precision computer-controlled calibration generator. The amplitude of its
 signal can be set to any value from 0 to 1 V with a resolution of 1/4096. The
 equivalent electronics noise of individual calorimeter channel lies within the
 range of $150\div350$ keV.

 To produce signals for the first-level trigger the SHAs are collected into
 160 modules corresponding to 160 towers (4 crystals from three layers
 located one under another). The towers divide calorimeter into 20 sectors of
 $18^\circ$ in azimuthal direction and into 8 rings of $18^\circ$ in polar
 direction.

\subsection{Energy calibration}

\begin{figure}
\begin{center}
\epsfig{figure=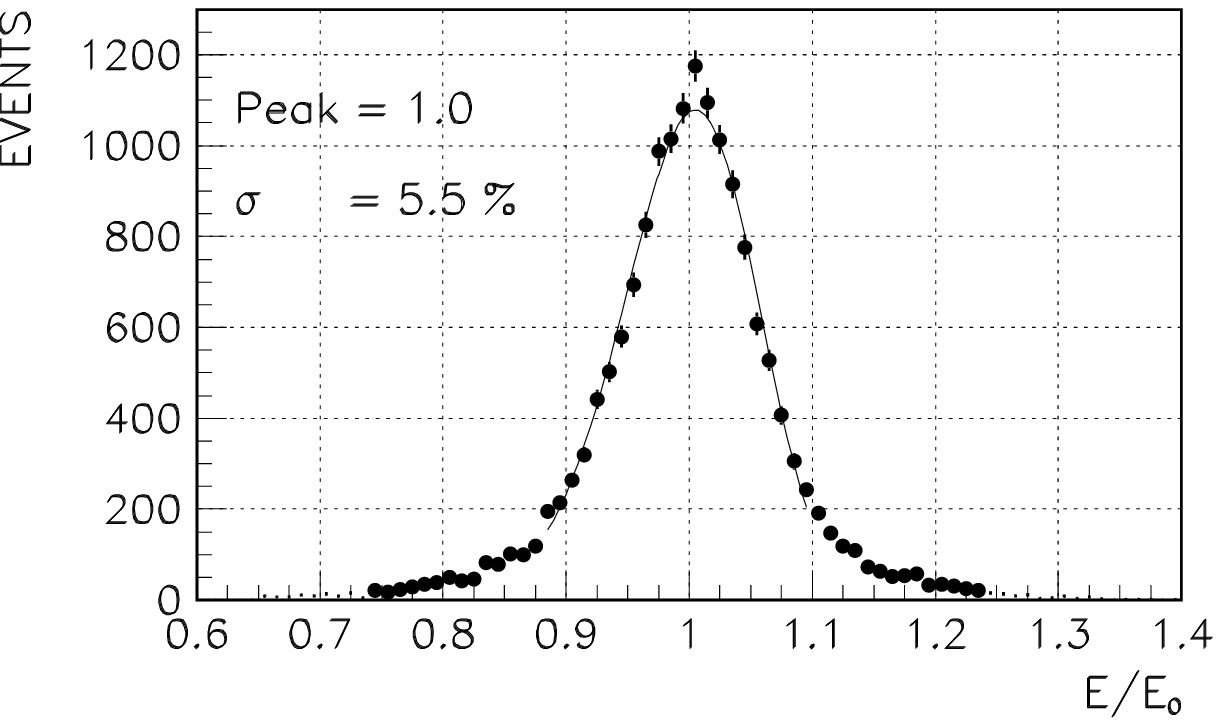,height=7.5cm}
\caption{Energy spectra for photons with energy $E_0 = 500$ MeV after cosmic
         calibration. $E$ is a measured energy.}
\label{ggcc}
\epsfig{figure=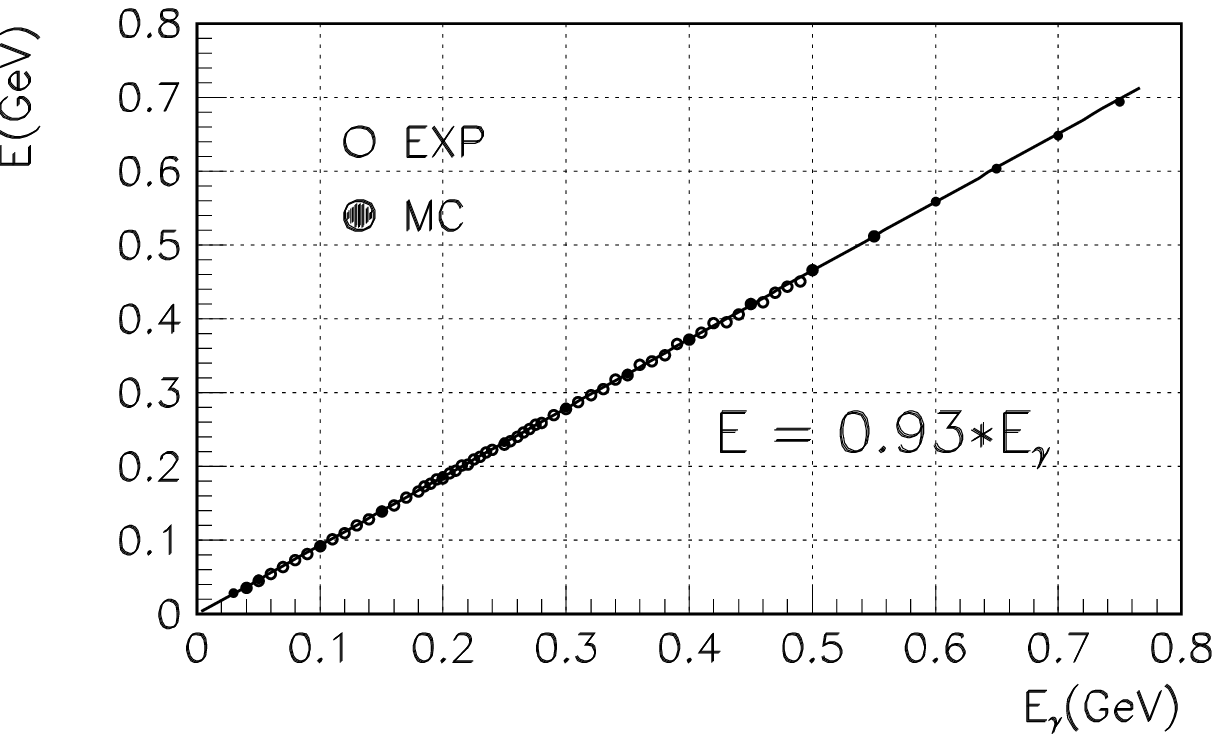,height=7.5cm}
\caption{Dependence of the most probable energy deposition on the photon
         energy.}
\label{gali}
\end{center}
\end{figure}

 Calorimeter is calibrated using cosmic muons \cite{ccc} and
 $e^+e^- \rightarrow e^+e^-$ process events \cite{ecc}.
 A fast preliminary calibration based on cosmic muons gives the constants for
 calculation of energy deposition in the calorimeter crystals. These constants
 are used for levelling responses of all crystals in order to obtain an uniform
 first-level trigger energy threshold over the whole calorimeter and
 represent seed values for the more precise calibration procedure,
 using $e^+e^- \rightarrow e^+e^-$ events. The cosmic calibration procedure
 is based on the comparison of the experimental and simulated
 energy depositions  in the calorimeter crystals for cosmic muons.
 The
 detailed description of the method is given in  \cite{ccc}. The statistical
 accuracy of $1\%$ in the calibration coefficients was achieved.
 After cosmic calibration the peak positions in the measured energy
 spectra for photons and electrons agree at a level of about $1\%$ with the
 actual particle energies (Fig.~\ref{ggcc}).

 The cosmic calibration procedure performed weekly, between the experimental
 runs takes less than 5 hours. For a one-week period between consecutive
 calibrations the coefficients stability was better than $1.5\%$. Changes in
 the conversion constants were also monitored daily, using the calibration
 generator. They could either drift slowly with electronics gain changes or
 show large leaps due to replacement of broken electronics modules.

 To achieve highest possible energy resolution, the precise OFF-LINE
 calibration procedure, based on analysis of $e^+e^- \rightarrow e^+e^-$
 events, was implemented.
 The calibration coefficients are obtained by the minimization
 of the r.m.s. of the total energy deposition  spectrum for
 the electrons with fixed
 energy. The detailed description of the procedure is given in \cite{ecc}. To
 obtain the statistical accuracy of about $2\%$ the sample of
 at least 150 electrons per crystal is needed. SND acquires such a sample
 daily when VEPP-2M operates in the center-of-mass energy $2E_0 \sim 1$ GeV.
 The average difference in calibration coefficients obtained using cosmic
 and $e^+e^-$ calibration procedures is about $4\%$, while the $e^+e^-$ 
 calibration improves the energy by $10\%$. For example, the energy
 resolution for 500 MeV photons improves from $5.5\%$ to $5\%$.

\subsection{Energy and angular resolution}

 The calorimeter energy resolution is determined mainly by the fluctuations
 of the energy losses in the passive material before ($0.17 X_0$) and inside
 ($0.17 X_0$) the calorimeter and leakage of shower energy through the
 calorimeter. The most probable value of the energy deposition for photons in
 the calorimeter is about $93\%$ of their energy (Fig.~\ref{gali}). 

 In order to compensate for the shower energy losses in passive material
 and improve energy resolution the photon energy is calculated as:
\begin{equation}
 E = \alpha_1 \cdot E_1 + \alpha_2 \cdot E_2 + \alpha_3 \cdot E_3, \label{eq:x}
\end{equation}
 where $E_1$ is energy deposition in the first and second layers of
 the central tower of the shower, $E_2$ is energy deposition in the first
 two layers outside the central tower, $E_3$ is energy deposition in the third
 layer, $\alpha_i$ are energy dependent coefficients. Here the tower is the
 three counters of the 1, 2 and 3 layers with the same $\theta$ and $\phi$
 coordinates and the central tower corresponds to the shower center of gravity.

 The $\alpha_i$ coefficients were determined from simulation of
 photons with energies from 50 to 700 MeV. For each photon
 energy the objective function:
 \begin{equation}
 M = \sum\limits_{k}(E^* - E_k )^2,
 \end{equation}
 was minimized over $\alpha_i$.
 Here $E^*$ is a known photon energy,
 $E_k$ is the energy calculated using expression (\ref{eq:x}), $k$ is a photon
 number. The energy dependences of $\alpha_i$ were approximated by the smooth
 curves. The approximation was done separately for the showers starting
 in different layers. As a result the calorimeter resolution was improved by
 $10 \%$.

\begin{figure}
\begin{center}
\epsfig{figure=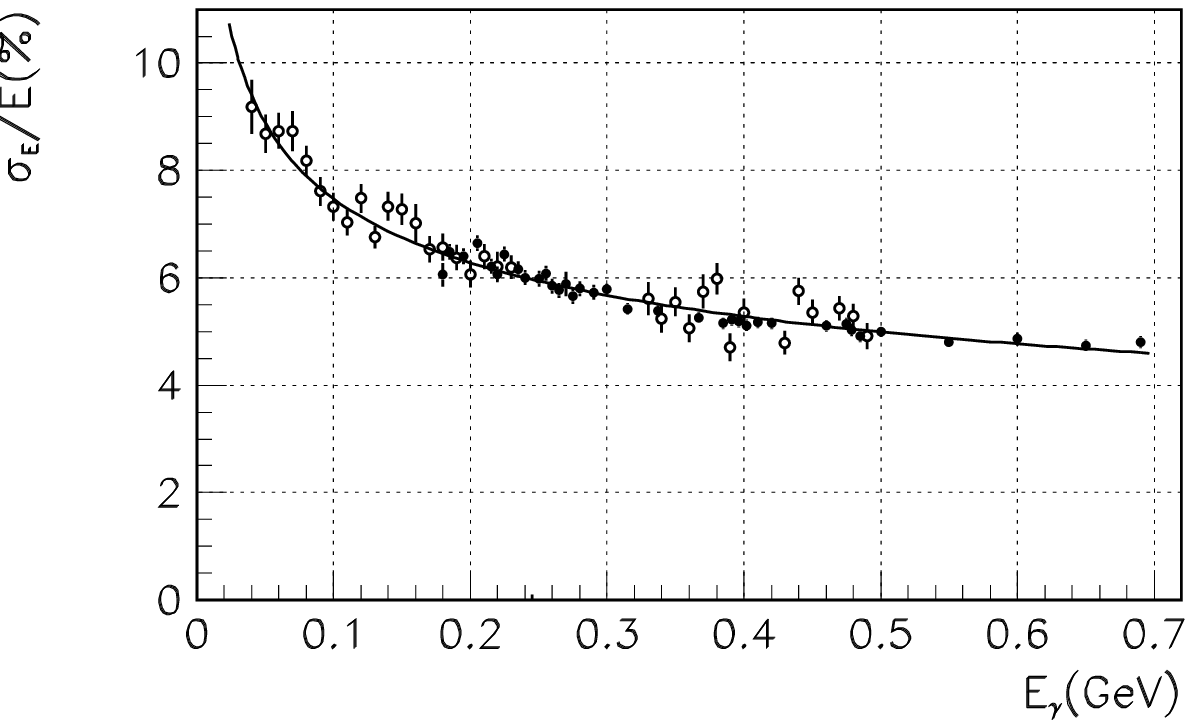,height=7.5cm}
\caption{Dependence of the calorimeter energy resolution on the photon energy,
         $\sigma_E/E(\%) = 4.2\% / \sqrt[4]{E(\mbox{~GeV})}$.
	 $E$ -- photon energy, $\sigma_E/E$ -- energy resolution obtained
	 using $e^+e^- \rightarrow \gamma \gamma$ (dots) and
	 $e^+e^- \rightarrow e^+e^- \gamma$ (circles) reactions.}
\label{resge}
\epsfig{figure=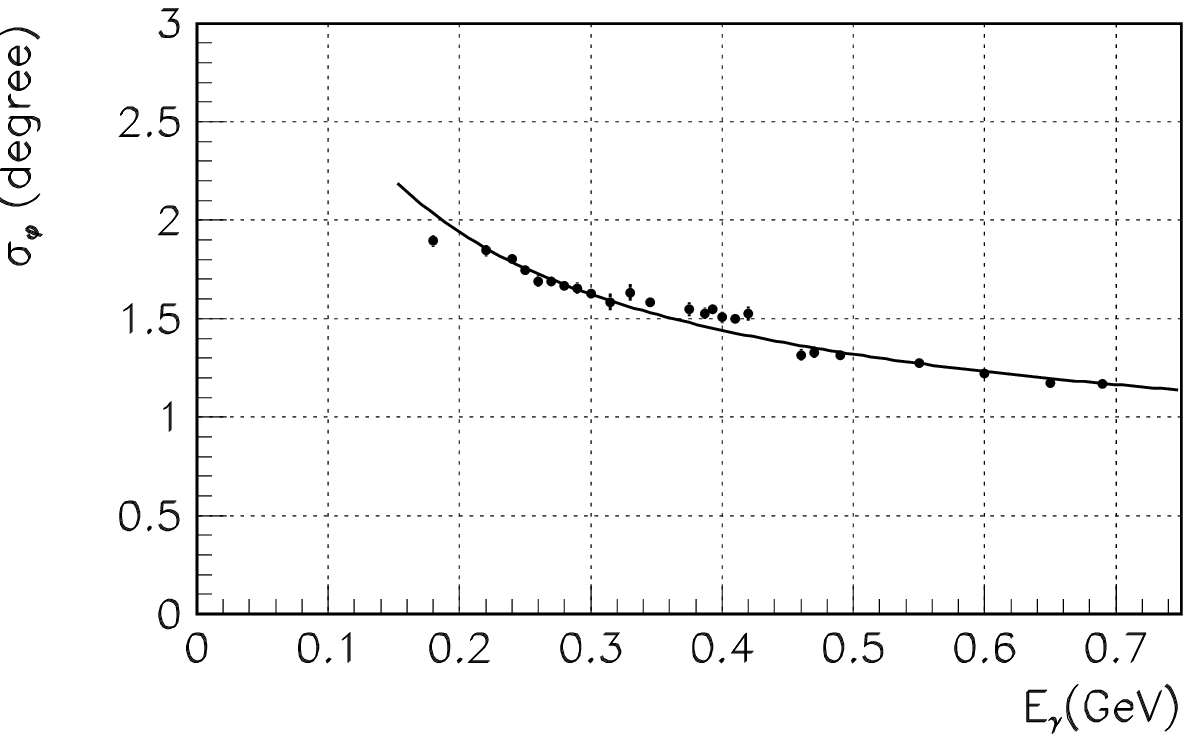,height=7.5cm}
\caption{Dependence of the angular resolution on the photon energy,
        $\sigma_\phi~=~0.82^\circ / \sqrt[]{E(\mbox{~GeV})} \oplus 0.63^\circ$.
        $E$ -- photon energy}
\label{resan}
\end{center}
\end{figure}

 The apparatus effects, such as nonuniformity of light collection
 efficiency over the crystal
 volume and electronics instability also affect energy resolution:
\begin{equation}
 \sigma_E/E(\%) = \sigma_1(E) \oplus \sigma_2(E) \oplus \sigma_3(E),
\end{equation}
 where $\sigma_1(E)$ is the energy resolution obtained using Monte Carlo
 simulation without effects mentioned above,
 $\sigma_2(E)$ is electronics instability and calibration accuracy
 contribution, $\sigma_3(E)$ is the contribution of nonuniformity of light
 collection over the crystal volume. For example, for 500 MeV
 photons $\sigma_E/E = 5\%$, $\sigma_1(E)=3\%$, $\sigma_2(E)=1.2\%$ and
 $\sigma_3(E)=3.8\%$. The dependence of the calorimeter energy resolution on
 photon energy (Fig.~\ref{resge}) can be approximated as:
\begin{equation}
 \sigma_E/E(\%) = {4.2\% \over \sqrt[4]{E(\mathrm{GeV})}}.
\end{equation}

 The distribution function of energy deposition in the SND calorimeter outside
 the cone with the angle $\theta$ around the shower direction was obtained
 using Monte Carlo simulation:
\begin{equation}
  E(\theta) = \alpha \cdot exp(-~\sqrt[]{\theta/\beta}),
\end{equation}
 It turned out that $\alpha$ and $\beta$ parameters are practically
 independent of photon energy in a wide energy range $50 \div 700$ MeV.
 The method  of estimation of photon angles based on this dependence was
 introduced in \cite{ivbe}. The dependence of the angular resolution on
 the photon energy shown in  Fig.~\ref{resan} can be approximated as:
\begin{equation}
 \sigma_\phi = {0.82^\circ \over \sqrt[]{E(\mathrm{GeV})}} \oplus 0.63^\circ.
\end{equation}

\begin{figure}
\begin{center}
\epsfig{figure=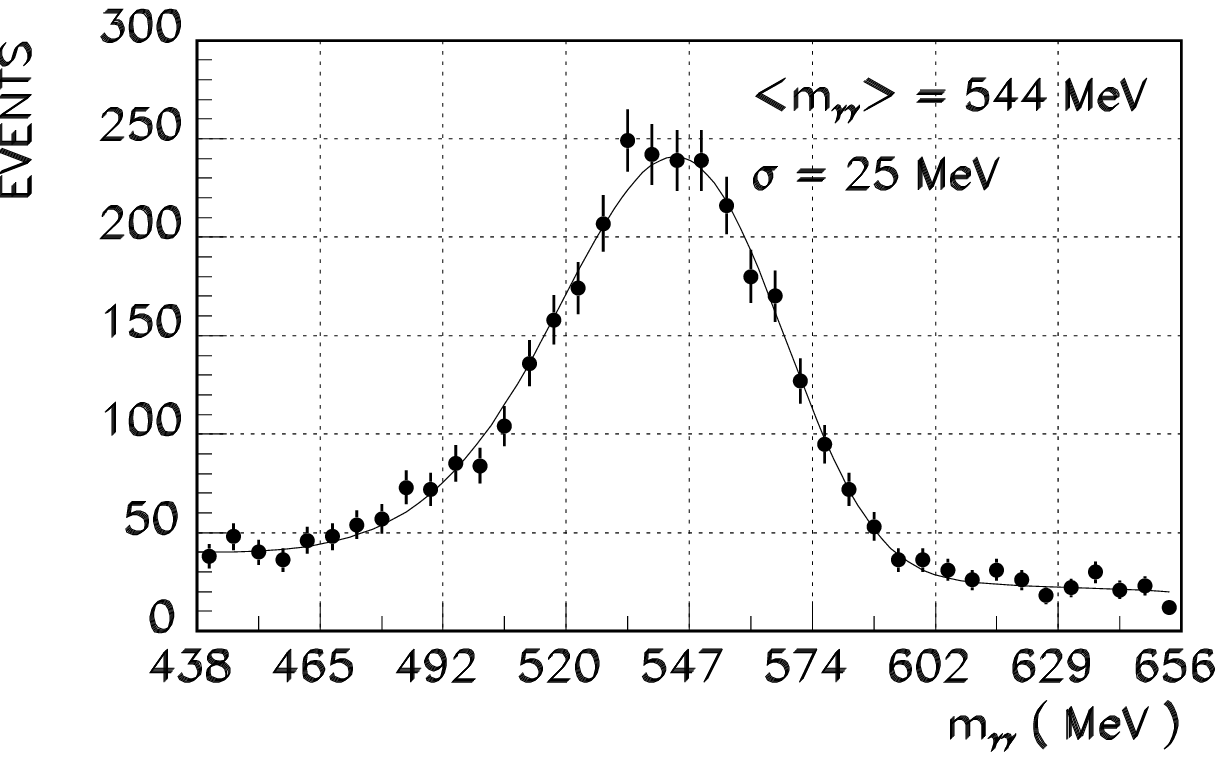,height=7.5cm}
\caption{Two-photon invariant mass distribution in experimental
         $\phi \rightarrow \eta \gamma$ events.}
\label{etag}
\epsfig{figure=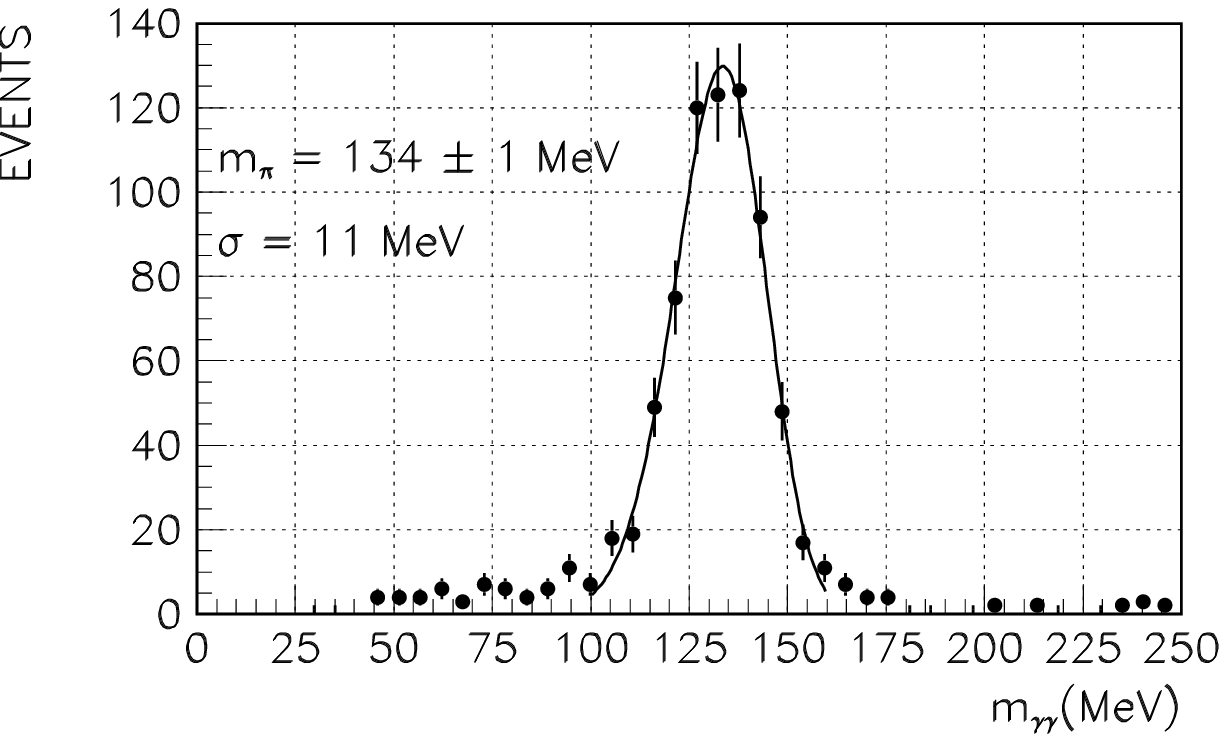,height=7.5cm}
\caption{Two-photon invariant mass distribution in experimental
         $\phi \rightarrow \pi^+ \pi^- \pi^0$ events.}
\label{pi3}
\end{center}
\end{figure}

\begin{figure}
\begin{center}
\epsfig{figure=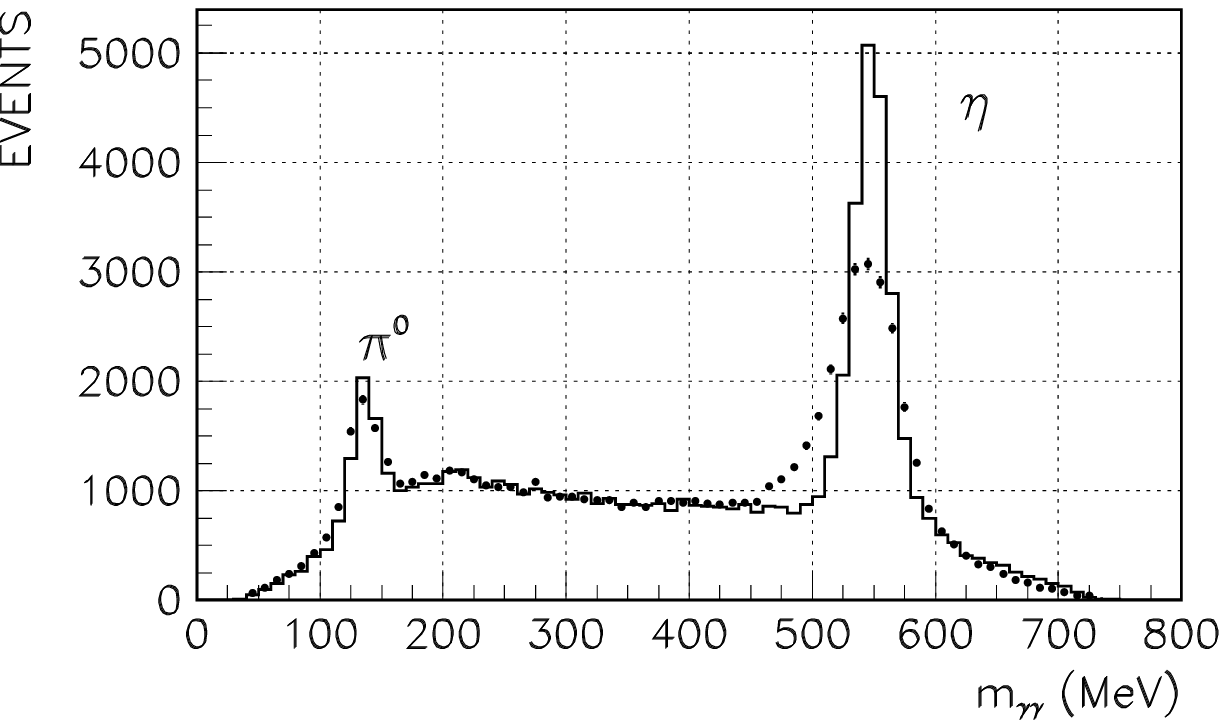,height=7.5cm}
\caption{Spectra of the invariant masses of photon pairs in
         $e^+e^- \rightarrow \gamma \gamma \gamma$ events before (dots) and
         after (line) kinematic fit.}
\label{ggg}		  
\epsfig{figure=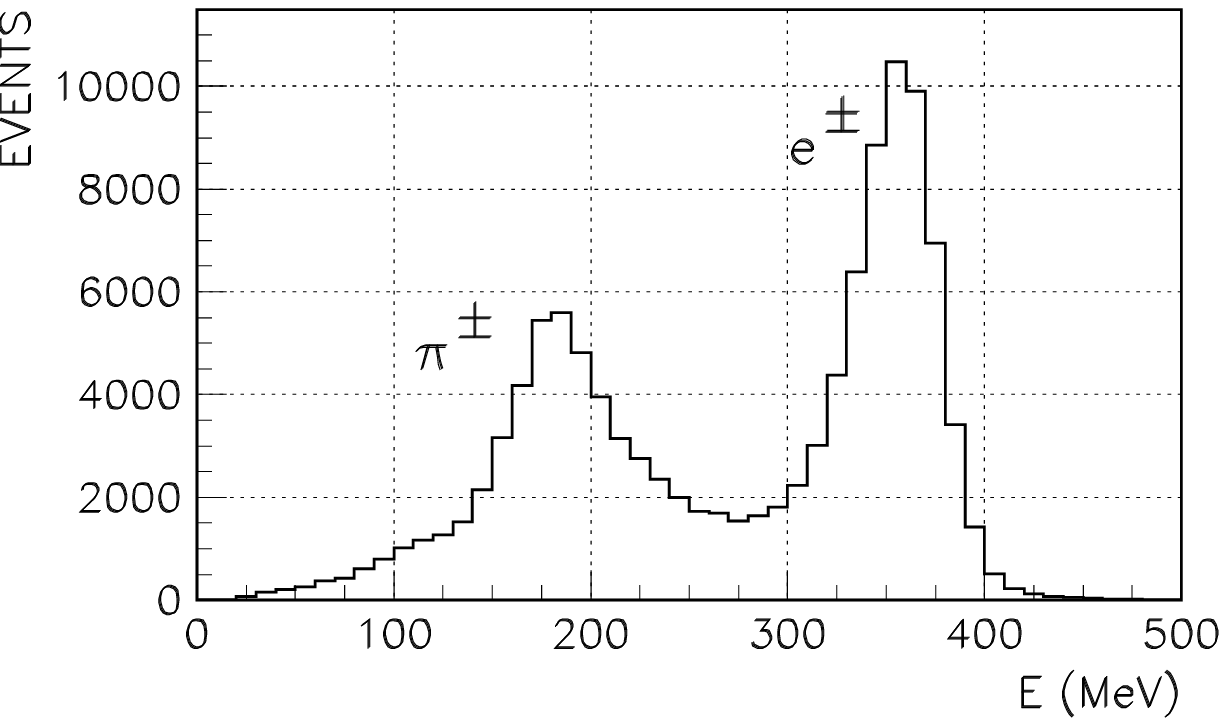,height=7.5cm}
\caption{Energy deposition spectra for 385 MeV $e^\pm$ and $\pi^\pm$}.
\label{sep1}
\end{center}
\end{figure}

 The two-photon invariant mass distributions in $\phi \rightarrow \eta \gamma$
 and $\phi \rightarrow \pi^+\pi^-\pi^0$ events (Figs.~\ref{etag},~\ref{pi3}),
 show clear peaks at $\pi^0$ and $\eta$ mesons masses. Invariant mass
 resolution is equal to 11 MeV for $\pi^0$ and 25 MeV for $\eta$.
 The kinematic fitting \cite{kin1} improving angular and energy resolution
 of the detector. For example for $\phi \rightarrow \eta \gamma, \pi^0 \gamma$
 decays kinematic fit improves two-photon invariant mass resolution by a
 factor of 1.5 (Fig.~\ref{ggg}).

\subsection{Particle identification}

\begin{figure}
\begin{center}
\epsfig{figure=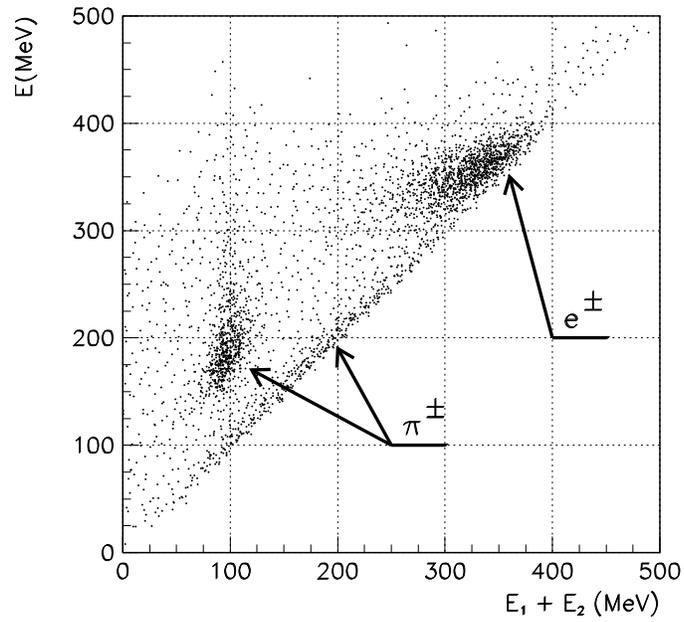,height=10cm}
\caption{Energy deposition in calorimeter first two layers vs total
         energy deposition for 385 MeV $e^\pm$ and $\pi^\pm$. $E$ -- total
         energy deposition, $E_1+E_2$ -- energy deposition in the first
         two layers}
\label{sep2}
\end{center}
\end{figure}

 The discrimination between electromagnetic and hadronic showers in the
 calorimeter is usually based on difference in total energy deposition
 (Fig.~\ref{sep1}). Multilayer structure of the SND calorimeter provides
 additional means of particle identification based on differences in
 longitudinal energy deposition profiles. The distributions of energy
 deposition over layers for $e^\pm$ and $\pi^\pm$ is shown in Fig.~\ref{sep2}.
 Two areas of pions concentration in this scatter plot correspond to nuclear
 interaction of pions and pure ionization losses in the first two layers of
 NaI(Tl). Utilizing differences in energy depositions for electrons
 and pions the special discrimination parameter was constructed. In the
 $\rho(770)$ energy region it provides $99\%$ selection efficiency for
 $e^+e^- \rightarrow \pi^+\pi^-$, keeping contamination by
 $e^+e^- \rightarrow e^+e^-$ events $\sim 1\%$

 The distribution of energy depositions in the calorimeter layers is also
 used for $K_L$ identification, for example in $\phi \rightarrow K_S K_L$
 decays, making feasible studies of rare $K_S$ decays.

 The $e/\pi$ and $\gamma/K_L$ separation parameters based on the differences in
 energy deposition profiles in transverse direction were also constructed.
 Their detailed description is given in \cite{ivbo}.

\section{Tracking system}

\subsection{Tracking system layout}

\begin{figure}[h]
\begin{center}
\epsfig{figure=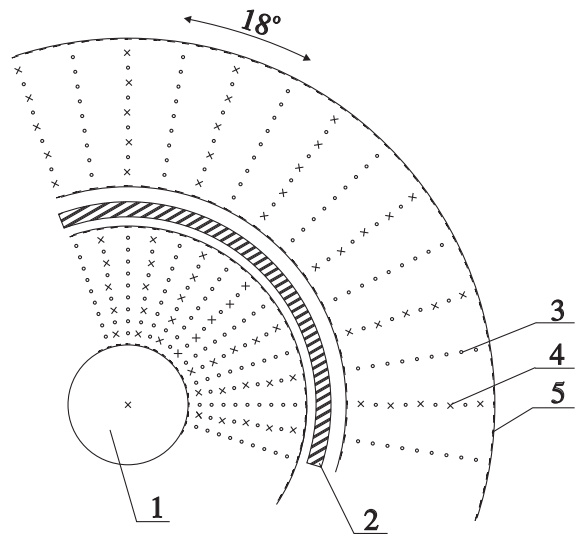,height=7cm}
\caption{SND coordinate system view across the beams: (1) beam pipe,
         (2) scintillation counter, (3) field shaping wires, (4) anode
         wires, (5) field shaping strips}
\label{dc}
\end{center}
\end{figure}

 The tracking system (Fig.~\ref{dc}) consists of two cylindrical drift chambers
 and a cylindrical plastic scintillation counter (CSC)\cite{cc} between them.
 
 The counter length, inner diameter and thickness are 37, 12.6 and 0.5 cm
 respectively. In azimuth direction it is divided into 5 segments.
 The counter has a wavelength shifter fiber readout. It provides the 
 time synchronization with beams collisions and produces signals for the
 first-level trigger. The time resolution of the counter is about 1.4 ns.
 
 The inner (``long'') drift chamber (LDC) has the length of 40 cm, outer and
 inner diameters of 4 and 12 cm respectively. The corresponding dimensions of
 the outer (``short'') drift chamber (SDC) are 25, 14 and 24
 cm. 

 Both chambers consist of 20 jet-type drift cells. Each cell has
 5 gold-plated tungsten sense wires with a diameter of 20 $\mu$m.
 A $\pm 300$ $\mu$m staggering of the sense wires in azimuth direction
 resolves the left/right ambiguity. Field wires are made of
 bronze-coated titanium with a diameter of 100 $\mu$m. The total numbers of
 sense and field wires in each chamber are 100 and 260 respectively.
 The drift chambers flanges are made of fiberglass with a thickness of 10 mm.
 The wires are fixed using copper pins, precisely positioned in the
 flanges. The cylindrical fiberglass walls of both chambers have three layers
 of copper electrodes: field strips, parallel to the wires, and sense
 (cathode) strips, transverse to the wires, with electrodes for signal output.
 The effective thickness of tracking system is about $0.1 X_0$.  Both chambers
 operate with a $90\%\mathrm{Ar}+10\%\mathrm{CO}_2$ gas mixture.

\begin{figure}[h]
\begin{center}
\epsfig{figure=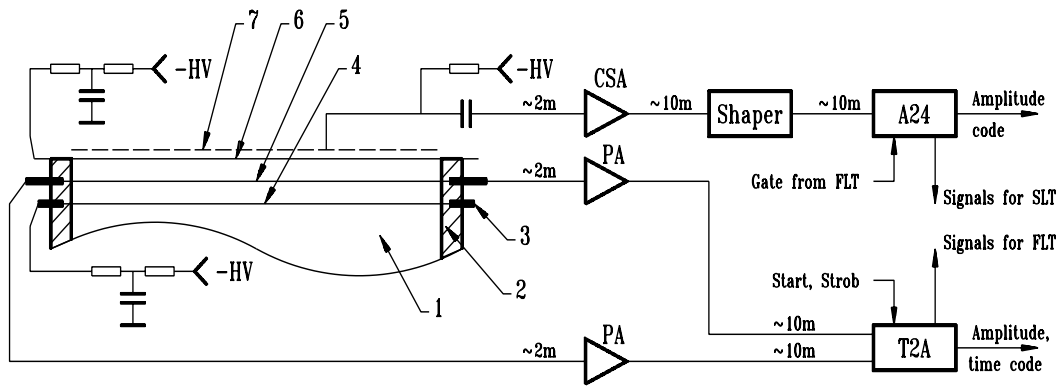,height=5cm,width=\textwidth}
\caption{Electronics channel of Drift chambers (DC): (1) DC gas volume,
         (2) end cap, (3) pins for wire attachment, (4) field shaping
         wires, (5) sensitive wires, (6) field shaping strips, (7) sensitive
         strips, (HV) potential from high voltage divider, (CSA) charge
         sensitive amplifier, (PA) preamplifier, (A24) 24 channel amplitude
         to digital converter, (T2A) time and two amplitude to digital
	 converters}
\label{dcel}
\end{center}
\end{figure}

 The circuitry of the electronics channel of the tracking system is shown in
 Fig.~\ref{dcel}. The field wires and strips operate at different potentials
 in the range of 1--3~kV to provide the uniform drift field. The sense wires
 operating at ground potential are DC coupled to preamplifiers, while for
 cathode strips capacitive AC coupling is used. The amplified analog signals
 from each sense wire are transmitted to the amplitude- and time-to-digital
 converters (ADC and TDC). So, two amplitudes from both ends and a common
 timing are measured for each wire.
 
\subsection{Calibration of the tracking system }

 The calibration of the tracking system consists of several procedures
 to obtain the coefficients for conversion of measured times and
 amplitudes into the track coordinates and specific ionization losses
 ($dE/dx$).

 The sense wires are calibrated electronically using the generator. The
 generator signals with varying amplitudes are switched between the
 preamplifiers on both ends of the wires. For each generator amplitude an
 average ADC count is calculated. These data provide calibration constants for
 $z$-coordinate measurement by means of charge division method:
\begin{equation}
 z = l_0 \cdot {{B_L Q_L - B_R Q_R} \over {Q_L + Q_R }}.
\end{equation}
 $Q_{L(R)} = A_{L(R)}U_{L(R)}$ is the charge collected at the left
 (right) end of the wire, $U_{L(R)}$ is the measured amplitude in ADC counts,
 $A_{L(R)}$ is the coefficient for conversion of the amplitude in ADC counts
 into units of electric charge, $B_{L(R)}$ is a conversion coefficient 
 from charge into a normalized coordinate, $l_0$ is the wire length.

 The $z$-coordinate is also measured by means of cathode strip readout. To
 obtain a conversion coefficients from the ADC counts into the input signal
 amplitudes, the electronics channels are calibrated using generator. The
 track coordinate can be calculated then as:
\begin{equation}
 z = Z_a+\Delta Z,
\end{equation}
 where $Z_a$ is the coordinate of the strip with the maximum amplitude
 (central strip), $\Delta Z$ is the correction obtained from the
 ratio of the amplitudes at all fired strips to the amplitude at the
 central strip.  $\Delta Z=f(q_{a-1},q_a,q_{a+1})$, where $a$ is the
 central strip number. Function $f$ was obtained using experimental cosmic
 muon sample, under assumption of their uniform distribution in $z$.

 The small differences in wire lengths $z$-coordinates of their centers 
 contribute to the accuracy of $z$-coordinate measurements by charge division.
 To eliminate this contribution a calibration procedure based on analysis of
 $e^+e^- \rightarrow e^+e^-$ events was performed. Cathode strips hit by one
 track were fitted and correction coefficients for each wire obtained. The
 corrected $z$-coordinate reads:
\begin{equation}
 z = z_w \cdot (1+\Delta L/L)-\Delta z_0,
\end{equation}
 where $z_w$ is the track coordinate measured by charge division, $\Delta L/L$
 is the corresponding wire length correction, $\Delta z_0$
 is the longitudinal bias of the wire with respect to the center of the
 tracking system.
 
 Due to absorption in the gas mixture the average signal amplitude on a wire
 decreases by 1.5 times per 1 cm of the drift distance. The dependence of the
 average amplitude on drift distance can be approximated as:
\begin{equation}
 A(R)=A_0(1-R/l),
\end{equation}
 where $R$ is a measured drift distance. Attenuation length $l$ is
 obtained using $e^+e^- \rightarrow e^+e^-$ events and
 $A_0$ is the calibration constant.

 For the conversion of the drift time into the distance $R(T)$
 from track to the wire the $r-\phi$ calibration based on uniformity of $\phi$
 distribution of $e^+e^-\to e^+e^- $ events was performed.
 
\subsection{Drift chambers efficiency and resolution}

 The efficiency of wires and cathode strips at low counting rates
 is close to $100 \%$. The drift chambers total counting rate is about 150 kHz
 for the beam currents $I_{e^+}=I_{e^-} \sim 50$ mA and luminosity of
 $\sim 2 \cdot 10^{30}$ $\mathrm{cm}^{-2} \mathrm{s}^{-1}$. The pile up
 hits degrade accuracy of coordinates measurements. In order to increase the
 accuracy the hits distant from the track, are removed from the
 fit. Thus the final wire efficiency in track reconstruction is
 about $95 \%$. The accuracies of coordinate measurements by drift times
 and charge division methods is about 180 $\mu$m and 3.6 mm respectively.

 The tracking system angular resolution was measured using distributions in
 $\Delta \theta$ and $\Delta \phi$ acollinearity angles between tracks in the
 $e^+e^- \rightarrow e^+e^-$ and $e^+e^- \rightarrow \pi^+\pi^-$ events
 as $\sigma_{\theta} = \sigma_{\Delta \theta} / \sqrt{2}$ and
 $\sigma_{\phi} = \sigma_{\Delta \phi} / \sqrt{2}$, where $\sigma$ is
 by definition the
 full width of the distribution at half-maximum divided by 2.36.

\begin{figure}[h]
\begin{center}
\epsfig{figure=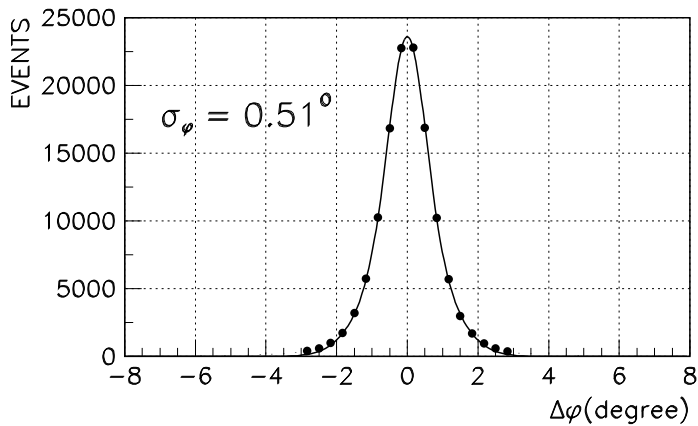,height=7.5cm}
\caption{The azimuthal acollinearity angle distribution for
         $e^+e^-~\rightarrow~e^+e^-$ events}
\label{phires}
\end{center}
\end{figure}

 The polar angle resolution $\sigma_{\theta}$ depends on the accuracy
 of $z$-coordinate measurement. The use of cathode strips together with
 sense wires provides 1.5 times improvement of the resolution relatively
 to charge division only. Finally, $\sigma_{\theta} = 1.7^\circ$ for electrons
 and $\sigma_{\theta} = 1.9^\circ$ for pions. The azimuth angle resolution
 $\sigma_{\phi} = 0.51^\circ$ for electrons (Fig.~\ref{phires}) and
 $\sigma_{\phi} = 0.54^\circ$ for pions. The impact parameter in $r-\phi$
 plane resolution $\sigma_R = 0.5$ mm. Small differences in angular
 resolutions for electrons and pions can be attributed to the different
 average ionization losses and $\theta$ distributions for these
 particles.

\begin{figure}[h]
\begin{center}
\epsfig{figure=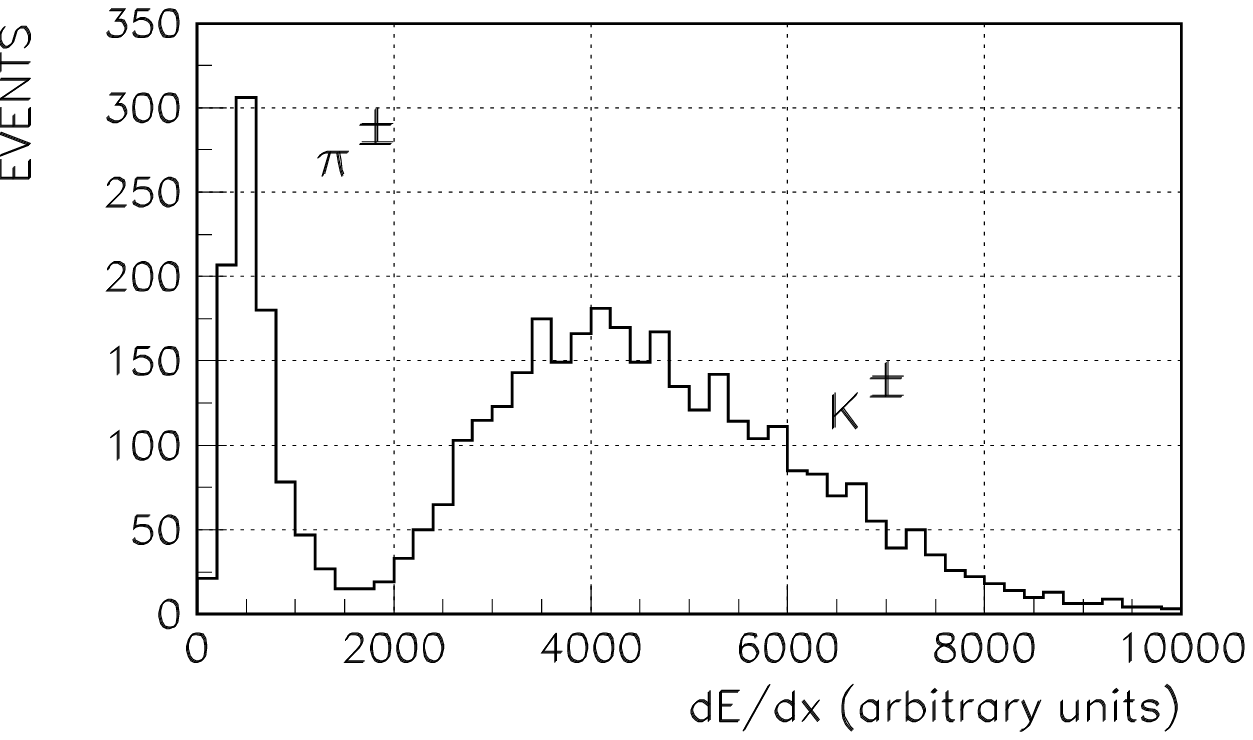,height=7.5cm}
\caption{$dE/dx$ distribution for events with charged particles in the
         center-of-mass energy range $2E_0 \sim 1$ GeV}
\label{dedx2}
\epsfig{figure=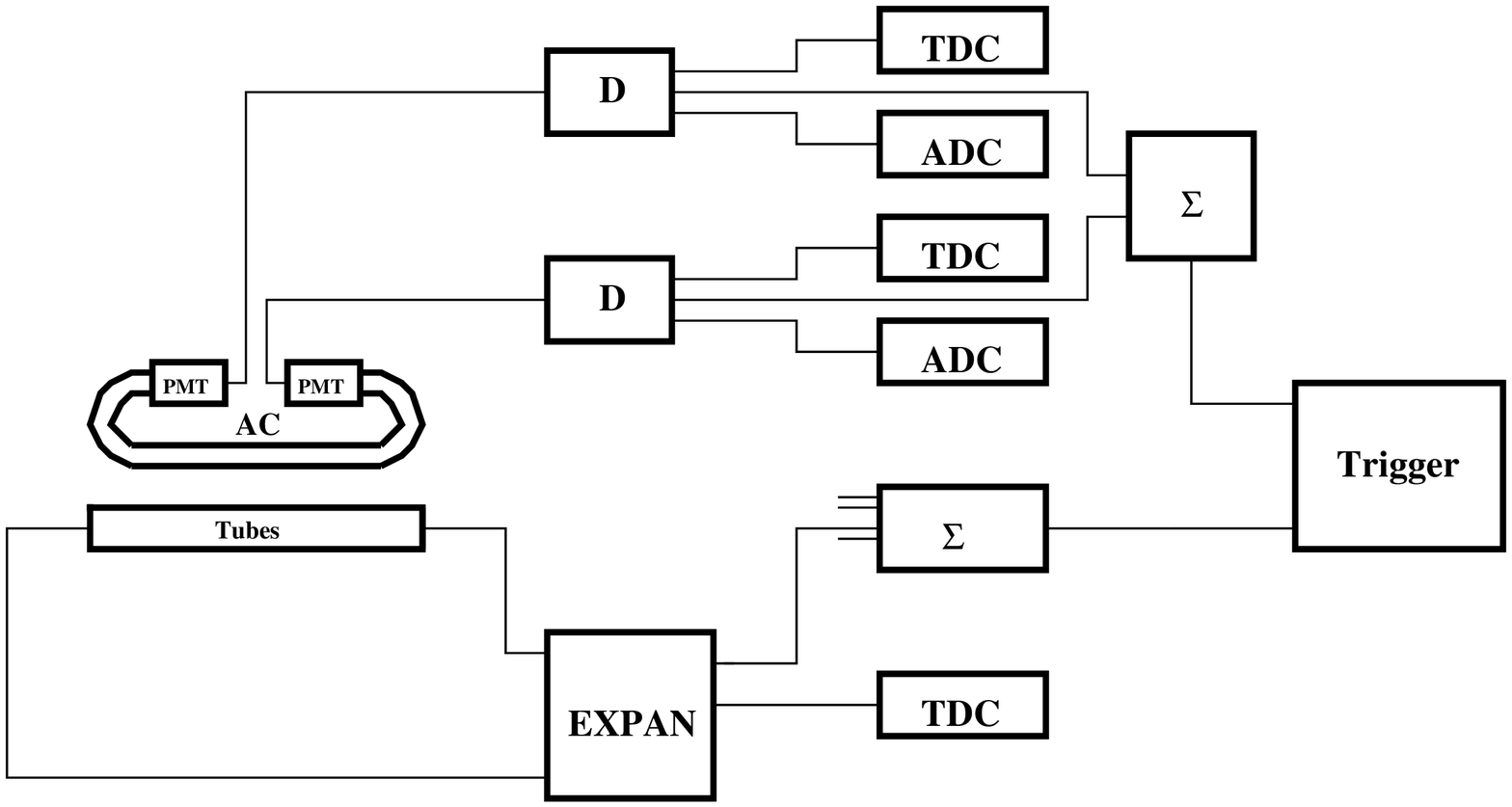,height=9cm,width=\textwidth}
\caption{Muon system electronics: (AC) scintillation counters,
         (Tubes) streamer tube, (D) discriminator, (EXPAN) time expander,
         (Trigger) first-level trigger, ($\Sigma$) logical summator,
	 (TDC) time to digital converter, (ADC) analog to digital converter}
\label{elout}
\end{center}
\end{figure}

 The achieved $dE/dx$ resolution is about $30 \%$ and it is not sufficient
 for $e/\pi$ separation, but $\pi^\pm/K^\pm$
 separation in the energy region of $\phi$-meson production is possible
 (Fig.~\ref{dedx2}).

\section{Muon system}
 
 The muon system provides suppression of cosmic background and identification 
 of the muons produced in the beams collisions at energies above
 450 MeV.
 
\subsection{Muon system layout}

 The muon system (Figs.~\ref{sndt},~\ref{sndf}) consists of 16
 ($200 \times 40 \times 1$ $\mathrm{cm}^3$) scintillation counters and two
 layers of streamer tubes. Each counter consists of two sheets of plastics
 scintillator glued together. Light is collected from two sides of the
 scintillator via the acrylic light guides to the pair of PM tubes with
 photocathode diameter of 40mm. Each counter is wrapped in aluminized mylar
 and put in a thin steel container.

 The barrel part of the streamer tubes system consists of 14 modules with
 40~cm width, two end-cap modules have the width of 80~cm. The module length
 is 2 m. Each module consists of 16 tubes arranged in two layers. The tubes
 are made of 300 $\mu$m stainless steel and have a diameter of 4 cm. Along
 the tube axis a 100 $\mu$m gold-plated molybdenum  wire is strung. Streamer
 tubes are filled with a $75\%\mathrm{Ar}+25\%n$-pentane gas mixture, which
 provides operation in a streamer mode at wire potential $\sim 4000$ V.

 The electronics channel of the muon system \cite{ustmu} is shown in
 Fig.~\ref{elout}. The analog signals from PMTs are carried to
 the discriminators where the amplitude and timing signals are formed, and then
 transmitted to ADC and TDC. The signals from opposite ends of the streamer
 tubes are applied to the time expander, where the signal with a duration
 proportional to their time difference is formed for further digitization
 by TDC. The signals from the tubes and counters are also
 used in the first-level trigger.

\subsection{Muon system calibration}

 $Z$-coordinate of the particle track is obtained using
 the time difference between the signals from the opposite ends of
 streamer tubes :
\begin{equation}
  Z = {l \over 2} - l \cdot {{C-\Delta T} \over C},
\end{equation}
 where $l$ is the wire effective length, known with the accuracy of
 about $2\%$,
 $\Delta T$ is the time difference between the signals from wire ends 
 expressed in TDC counts,
 $C$ is the calibration coefficient, obtained using $\Delta T$ distribution
 for cosmic muons. Stability of the electronics was checked daily 
 using calibration generator.
 
 To obtain the time of scintillation counter hit the following expression
 is used:
\begin{equation}
 t = K \cdot T + \Delta t,
\end{equation}
 where $T$ is a TDC count, $K$ is a coefficient for TDC count to time
 conversion, obtained from cosmic muons spectrum, $\Delta t$ is a correction
 accounting for differences in the individual cables lengths and PM tubes
 delay times. $\Delta t$ is obtained as the
 time difference between the hits in muon scintillation counter and
 tracking system scintillation counter, using $e^+e^- \rightarrow \mu^+\mu^-$
 events and cosmic muons. Cosmic muons are also used for the energy
 calibration of the scintillation counters (average energy losses are
 $2$ MeV/cm). The energy deposition in the counter $E = k \cdot U$, where $U$
 is the signal amplitude in ADC counts, $k$ is the coefficient of conversion
 from ADC counts to the units of MeV. To consider the amplitude dependence on
 the $Z$-coordinate of the muon track due to the light absorption in
 scintillator the muon track is fitted using streamer tubes.
 
\subsection{Efficiency and time resolution of the muon system}

 The muon system veto signal provide the 50 times cosmic background
 suppression in actual experimental conditions.

 A longitudinal coordinate of the track is obtained from the tube number. If
 only one tube is fired, then resolution $\sigma = d / \sqrt{12} = 1.2$ cm,
 where $d=4$ cm is the tube diameter. When the tubes in both layers are hit
 the resolution is $\sigma = 0.6$ cm determined by the shift between
 layers. The $Z$-coordinate resolution $\sigma_Z = 3.3$ cm.

\begin{figure}[h]
\begin{center}
\epsfig{figure=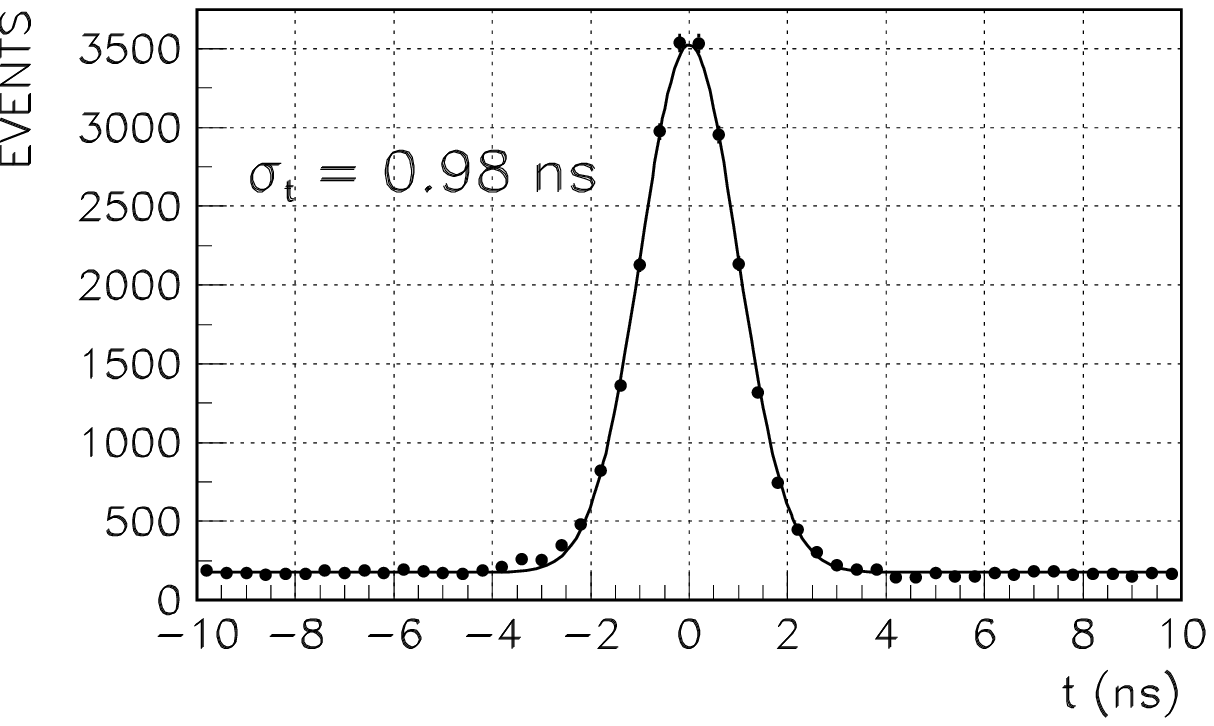,height=7.5cm}
\caption{Time distribution of scintillation counters signals with respect to
         the beams collision time in $e^+e^- \rightarrow \mu^+\mu^-$ events
         at $2E_0 \sim 1$ GeV. The cosmic muons background is also seen as a
	 uniform contribution.}
\label{tres}
\epsfig{figure=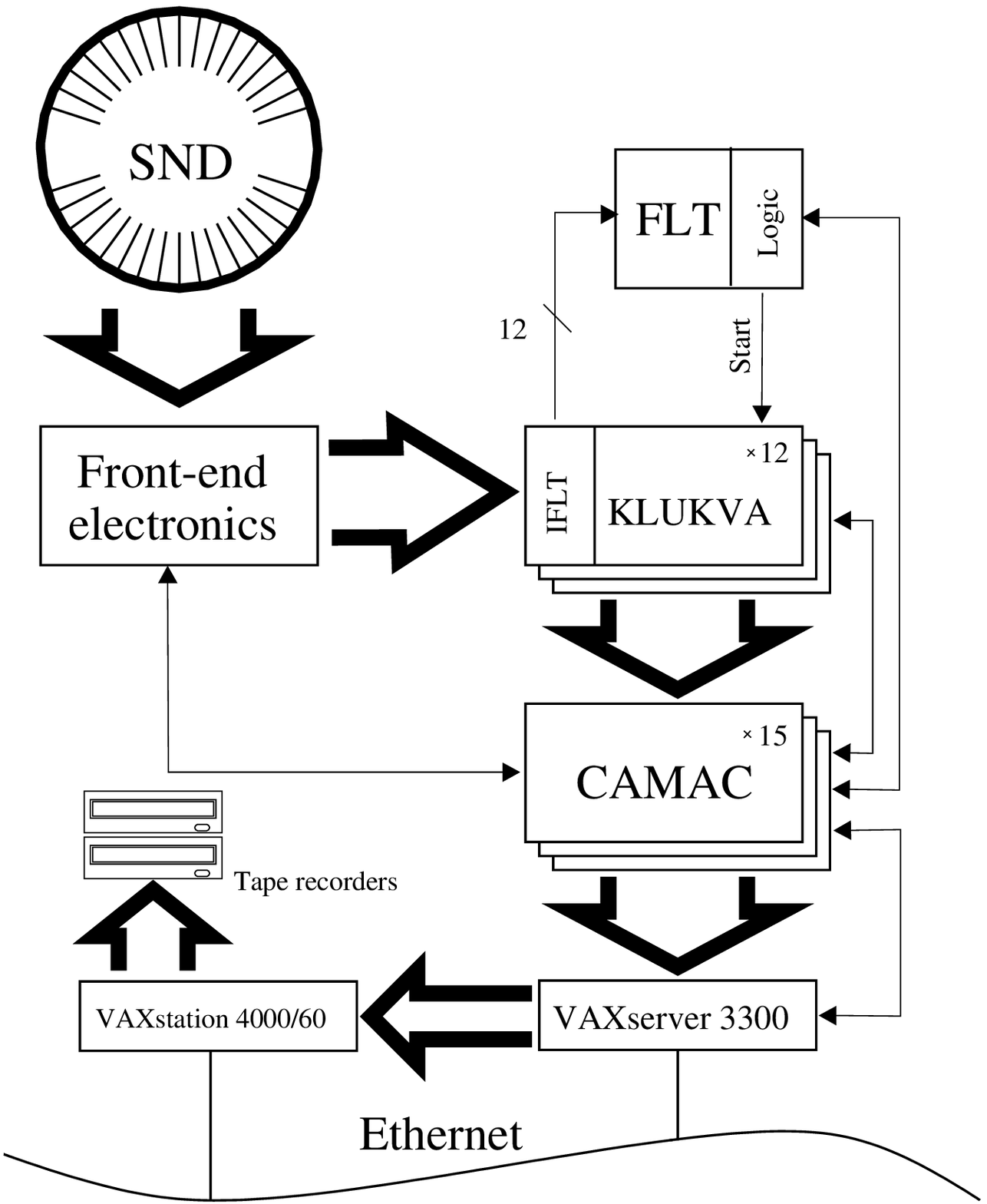,height=9cm}
\caption{SND data acquisition system: KLUKVA crates,
         (FLT) first-level trigger, (IFLT) FLT interface modules,
         (Logic) modules of FLT logic and decision, (CAMAC) CAMAC crates}
\label{daq}
\end{center}
\end{figure}

 To obtain the time of scintillation counters hit with respect to beams
 crossing, special corrections for light propagation time were applied.
 The corrections take into account the effective velocity of light in
 the scintillator ($\sim 15$ ns/cm) and dependence of the counter firing time
 on the signal amplitude due to fixed discriminator threshold. After
 corrections the time resolution of the counters is 1 ns (Fig.~\ref{tres}).
 
\section{Data acquisition system}

\subsection{Electronics}

 The SND data acquisition system  (DAQ) \cite{snddaq} (Fig.~\ref{daq}), 
 based on KLUKVA electronics standard \cite{klukva} developed at BINP.

 Analog signals from the detector elements come to the front-end electronics
 located on SND systems. Then the signals are carried via screened twisted
 pairs to digitizing electronics (ADCs and TDCs). The signals from the
 calorimeter and cathode strips are shaped by shaping amplifiers before
 digitization. The digitizers and shapers reside in KLUKVA crates. One crate
 holds 16 data conversion modules, one first-level trigger (FLT) interface
 module (IFLT) and readout processor module (RP). The data conversion modules
 provide data, which are transferred via KLUKVA crate bus to IFLT modules for
 further use in FLT.

 Two types of digitizers are used: TDCs with``common stop'' and
 ADCs with ``common start''. The first type \cite{t2a} provides the drift time
 measurement. Signals from sense wires start 250 MHz counters and
 FLT signal stops drift time digitizing. If FLT signal does not arrive within
 1 $\mu$s the counters reset. The second type
 \cite{a32} is used for digitization of signals from the calorimeter and 
 DC cathode strips. The digitization starts by FLT signal and lasts 80~$\mu$s.
 The  digitizers (about 3000 channels) occupy 13 KLUKVA crates.

 After the digitization the data are extracted from the ADC and TDC modules
 by the RP in 100~ns.
 An RP memory contains the pedestals of all channels placed
 in the crate. RP compares the data with pedestals and
 performs zeros suppression.
 As a result average event size decreases down to about 1kB. The RP internal
 memory keeps digitized data for two events, providing with the internal
 registers of ADCs and TDCs 3-level derandomizing of data flow. The time
 needed for event digitization and processing is about 200~$\mu$s. The data
 from RP are transmitted to the ON-LINE computer via special CAMAC modules
 \cite{klukva}.

\subsection{First-level trigger}
 
\begin{figure}[h]
\begin{center}
\epsfig{figure=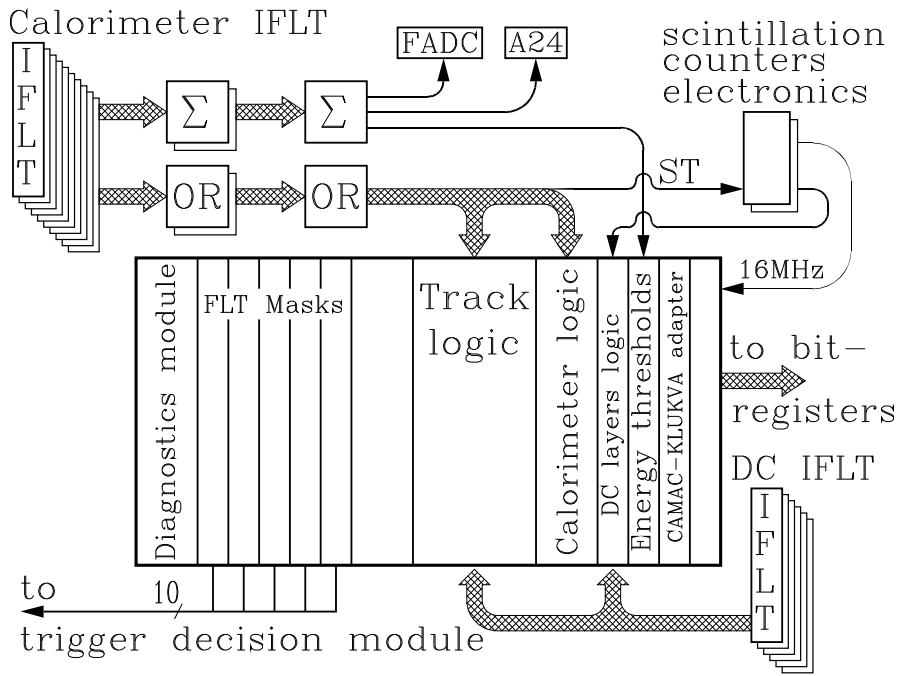,height=9cm}
\caption{First-level trigger (FLT) structure scheme: (IFLT) FLT interface
         modules, (FADC) ADC ``total energy deposition'', (A24) ADC,
         ($\Sigma$) -- logical summator}
\label{trig}
\end{center}
\end{figure}

 The fast analog signals of energy deposition in calorimeter towers and logical
 signals from wire hits in drift chambers are collected by IFLT modules.
 The IFLTs and
 analog and logical summation modules form the signals for specialized
 trigger modules:
 Calorimeter Logic, Track Logic, Layers Logic, and Energy Thresholds, producing
 48 different FLT components (Fig.~\ref{trig}).
 The input information for Calorimeter Logic \cite{logcal} is 20 signals
 from sectors and 8 signals from rings, produced by the logical
 summation of the hit towers with the energy deposition higher than 25 MeV.
 The sectors signals go to the address bus of 1MB static RAM, where the
 look-up table is stored. The information from rings from another
 32 kB RAM is added. So, according to the RAM contents, the components
 of the FLT depending on the number of hit towers and their relative
 positions are produced. FLT components can be easily changed
 by RAM reprogramming.

 The Energy Thresholds module contains 10 discriminators with programmable
 threshold. As an input information the analog signals of total energy
 deposition and energy deposition at ``large'' angles are used.

 A Track Logic \cite{logcam} is constructed using the programmable logic
 devices. As the input information the following logical signals are used: 100
 signal lines from LDC wires, 20 signals from the first two layers of SDC and
 20 signals from the calorimeter sectors.
 Hit wires patterns in LDC corresponding
 to tracks from interaction point are looked for. Then the track
 continuation in SDC and calorimeter is checked. Resulting signals contain
 information on the number of found tracks
 and their relative positions. The Layer Logic is based on static
 RAM and as input data uses 10 logical OR signals from LDC and SDC layers.
 It produces three components, each being the various combination of
 input signals, according to the RAM contents.
  
 The FLT logic is implemented as a  pipeline working at 16~MHz clock rate,
 which is the collider beam circulation frequency. Such FLT organization
 provides zero ``dead'' time, while the FLT decision latency is about 800 ns.

 The main source of background are electromagnetic showers produced in
 the collider magnets and lenses by stray particles from the beam. The FLT
 background suppression is based on the following differences in the properties
 of physical and background events:
\begin{enumerate}
\item
 the total energy deposition in the calorimeter for physical events is larger
 than for background events;
\item
 hits in physical events form two or more compact clusters in the  calorimeter;
\item
 particles in physical events originate from the collider interaction 
 region;
\item
 in the background events the main part of the total energy deposition in the
 calorimeter is localized at ``small'' angles. 
\end{enumerate}
 The most often used FLT components are:
\begin{enumerate}
\item
 ``OR'' of all calorimeter towers used as a master signal for FLT Mask
 Modules with a time resolution better 
 than 5 ns, while the time between collisions is
 60 ns;
\item 
 two well separated towers hit;
\item
 two collinear towers hit;
\item
 hit tower at an angle larger than $54^\circ$ with respect to the beam;
\item
 two hit towers with azimuth angle between them larger than $90$ degrees;
\item
 the total energy deposition and the energy deposition at ``large'' angle is
 larger than certain threshold;
\item
 LDC track, LDC track continued in SDC, and LDC track continued
 in SDC and calorimeter;
\item
 two separate tracks in LDC;
\item
 the muon system veto.
\end{enumerate}

 Final FLT decision is produced by 10 Mask Modules, each taking 48
 FLT components as inputs. It is possible
 to define 10 different trigger types by writing appropriate masks into the
 Mask Modules. Four trigger types select events containing charged
 particles (``charged'' triggers),
 other four select events with neutral
 particles only (``neutral'' triggers), one trigger selects
 cosmic muons for muon system calibration and the last selects the
 $e^+e^- \rightarrow e^+e^-$ events for on-line luminosity measurement.
 For all ``neutral'' triggers the muon system veto is enabled.

 In average experimental conditions the calorimeter and drift chambers
 counting rates are about
 80 kHz and 150 kHz respectively. The FLT
 reduces the trigger rate to a level of $\sim 80$ Hz.

\subsection{Data Taking}

 Data from RP RAMs are read into the VAXserver 3300 computer via the CAMAC
 interface modules at a rate of 2 ms/event.
 The computer RAM is used for data flow derandomization. Then the data are
 transmitted to the VAXstation-4000/60 workstation, where events are packed in
 a special SND format \cite{cocha}. For packed events the partial
 reconstruction (see section 7) is performed. The reconstructed events are
 processed by the software 3-rd level trigger which
 suppresses beam background and
 cosmic events, decreasing the event rate by a factor of 2. In addition it
 counts $e^+e^- \rightarrow e^+e^-$ and $\gamma \gamma$
 events for on-line luminosity monitoring. The total time required for partial
 reconstruction and 3-rd level
 trigger decision is $\sim 10$ ms/event. Average DAQ dead time
 is about $10\%$.

 Events selected by the software trigger are written on  file
 on the local disk. After each experimental run files are automatically
 saved to 4 GB tapes by the data management system ART \cite{snddaq,art}.

 During the run several concurrent processes collect information about
 detector subsystems, DAQ, and collider performance. Real-time access to these
 data is provided by information management system IMAN \cite{iman}.
 The special process monitors this information and rises alarm if hardware
 or software problems are detected.
 
 Stability of SND subsystems is monitored using precision generator signals
 to test all electronics channels. This procedure takes about 20 minutes
 and is performed daily.
 
\section{Data processing}

 Large volume of data collected during experimental runs requires quick and
 adaptable processing. The data processing is based on CERN software such as
 HBOOK \cite{hb}, HIGZ \cite{hg}, PAW \cite{paw}, MINUIT \cite{min}, as well
 as programs developed in BINP:
\begin{enumerate}
\item 
 GIST \cite{gist} provides common environment for data processing, user
 interface with other codes, data storage in format compatible with HBOOK
 package for further analysis using PAW;
\item
 COCHA \cite{cocha} provides the data input/output and storage. The data
 representation is based on entity -- relationship model;
\item
 event reconstruction code;
\item
 code for events visualization based on HIGZ graphics package;
\item
 ART \cite{art} provides transparent access to data stored on tapes and disks.
\end{enumerate} 
 
 Processing of experimental data consists of several steps:
\begin{enumerate}
\item
 SND subsystems are calibrated using $e^+e^- \rightarrow e^+e^-$ and
 $\mu^+\mu^-$ events;
\item
 events are reconstructed and parameters of particles are calculated;
 according to  these parameters the events divided into $\sim 20$ different
 physical classes;
\item
 for each experimental run the $e^+e^- \rightarrow e^+e^-$ and
 $\gamma \gamma$ events are counted for the integrated luminosity
 determination, efficiencies of SND systems are obtained; collected
 information about SND and collider performance is used in the analysis,
 for example, to determine the beam energy more precisely;
\item 
 the selected events are used for the physical analysis.
\end{enumerate}

 Event reconstruction is performed in a following sequence. First 
 step is a search
 for separated clusters in the calorimeter.
 Then the track reconstruction in drift
 chambers is performed. Tracks are linked to the calorimeter clusters. The
 clusters with energy depositions of more than 20 MeV not linked to
 tracks in DC are considered as photons. For
 charged tracks no requirements on
 energy depositions in the calorimeter are imposed. Finally parameters of
 reconstructed particles are calculated: angles, energies, $dE/dx$ losses.
 
\begin{figure}
\begin{center}
\epsfig{figure=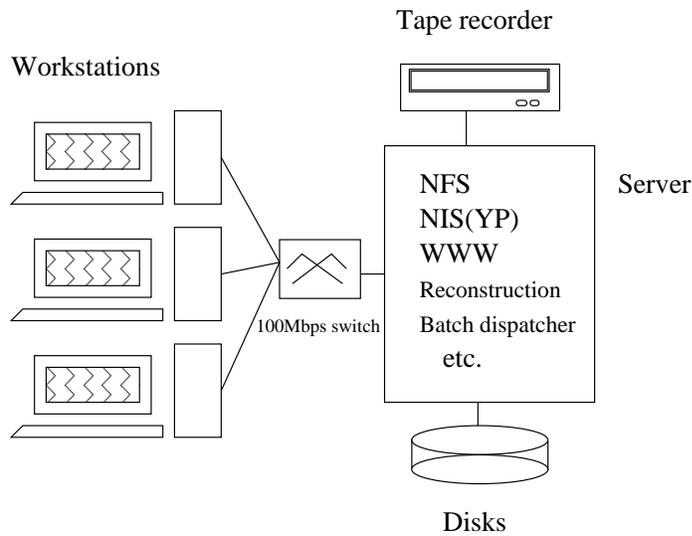,height=7cm}
\caption{The computing system.}
\label{clus}
\end{center}
\end{figure}

 The computer system (Fig.~\ref{clus}) used for reconstruction and analysis
 of experimental data includes several desktop Pentium PCs running
 Linux 2.0 and a Sun Enterprise 450 server under Solaris 2.6 operating system.
 All computers are connected via 100 Mb Ethernet to a local network.
 
 The server provides common software and user disk space for desktop
 PCs, along with batch system and network services. It also runs event
 reconstruction code at an average speed of about 150 events per second.
 
\section{Monte Carlo simulation}

 The Monte Carlo simulation of the SND is based on UNIMOD \cite{unimod}
 package.
 SND geometrical model comprises about 10000 distinct volumes. The  primary
 generated particles are tracked through the detector media taking into
 account the following effects: ionization losses, multiple scattering,
 bremsstrahlung of electrons and positrons, Compton effect, $e^+e^-$ pair
 production by photons, photo-effect, unstable particles decays, interaction
 of stopped particles, nuclear interaction of hadrons.
 After that the signals produced in each detector element are simulated.
 To provide the adaptable account of experimental conditions: the electronics
 noise, signals pile up, the actual time and amplitude resolution of
 electronics channels. Broken channels are also taken into account during
 reconstruction of Monte Carlo events.
 
\begin{figure}[h]
\begin{center}
\epsfig{figure=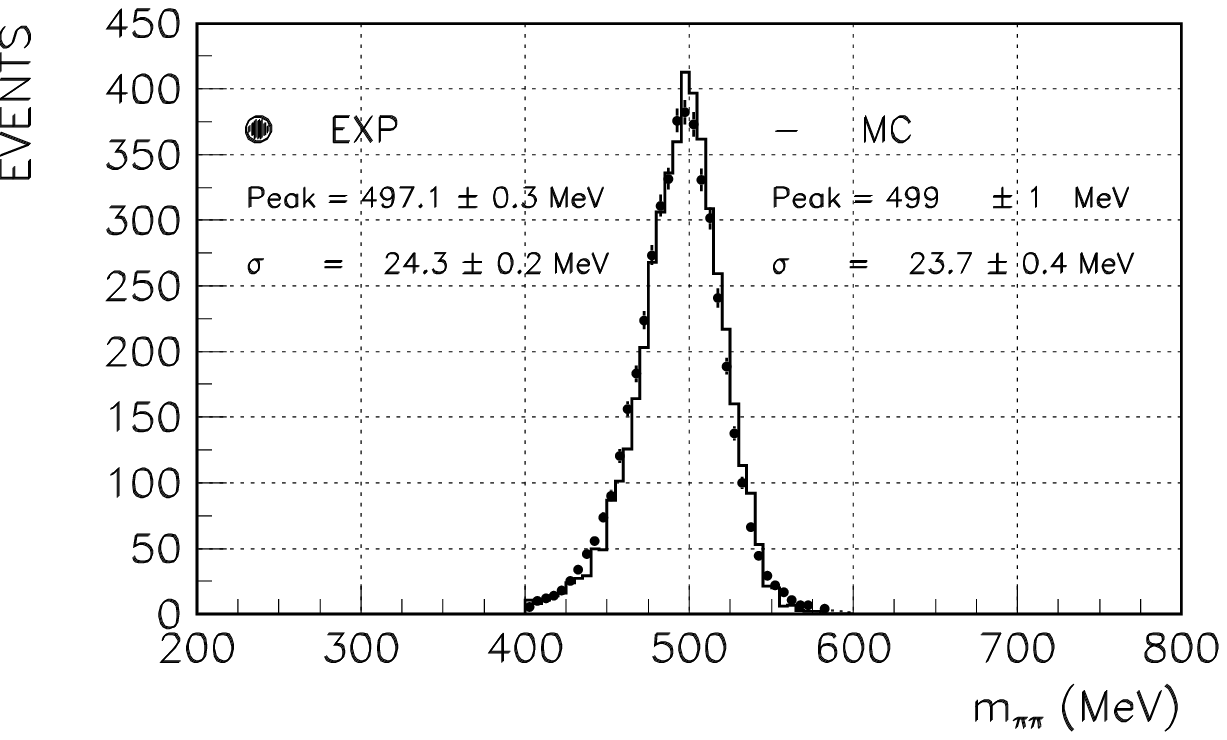,height=8cm}
\caption{Invariant mass distribution of $\pi^0$-mesons pairs from the
         $K_S \rightarrow \pi^0 \pi^0$ decay.}
\label{ksklres}
\epsfig{figure=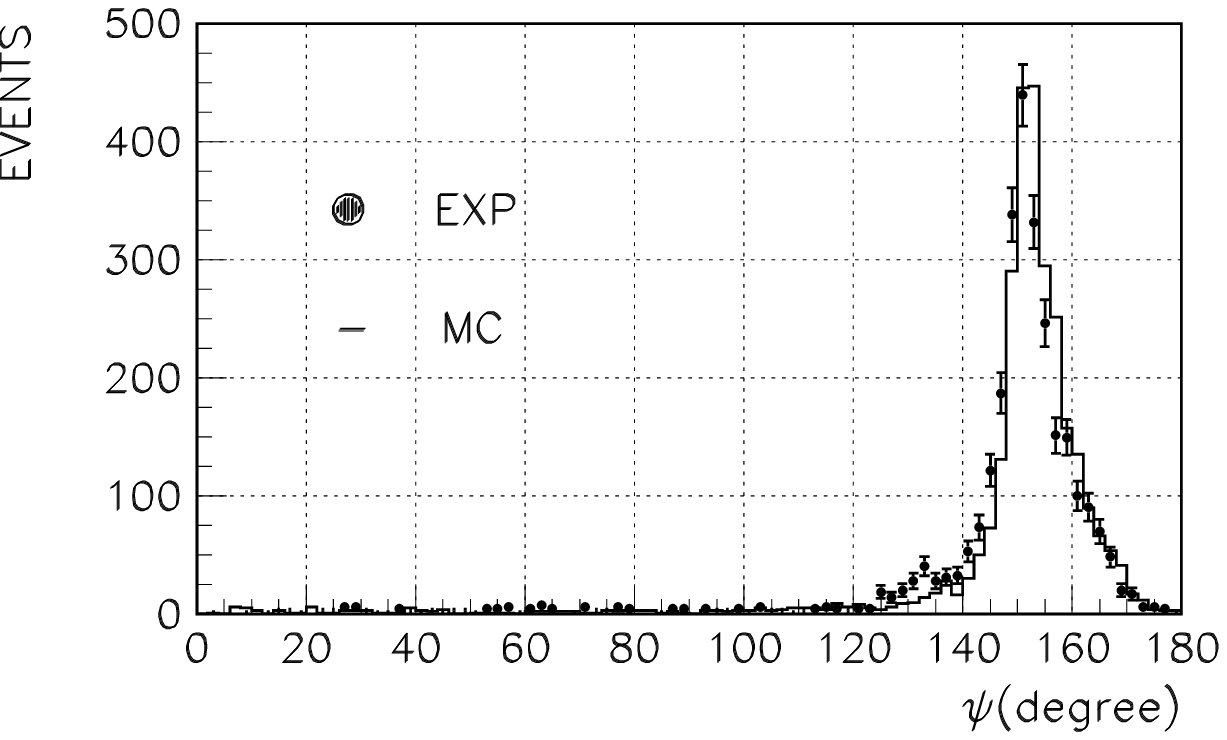,height=8cm}
\caption{Angle $\psi$ between pion pairs from the $K_S \rightarrow \pi^+ \pi^-$
         decay}
\label{ks2p}
\end{center}
\end{figure}

 Comparison of experimental and simulated distributions are shown
 in Figs.\ref{ksklres},~\ref{ks2p}.

\section{Conclusions}

 The design and performance of the SND detector operating at VEPP-2M $e^+e^-$
 collider since 1995 is described. During this time the total integrated
 luminosity of $\sim~25$~$pb^{-1}$ in the center-of-mass energy range
 $2E_0 = 0.4 \div 1.4$ GeV was accumulated. New rare radiative decays
 $\phi \rightarrow \pi^0 \pi^0 \gamma$ \cite{f0g}, $\eta \pi^0 \gamma$
 \cite{a0g} and OZI and G-parity suppressed decay
 $\phi \rightarrow \omega \pi^0$ \cite{ompi0} were observed.
 The $\phi \rightarrow \eta^\prime \gamma$ \cite{etg} decay existence was
 confirmed. Many other decays and hadronic production cross sections were
 measured with higher accuracy. More integrated luminosity is to be accumulated
 in the region of $\rho$ and $\omega$ mesons providing data for studies of
 their rare decays.

\section{Acknowlegement}

 This work is supported in part by STP `Integration'', grant No.274 and Russian
 Fund for Basic Researches, grant No.96-15-96327.

\newpage

\end{document}